\documentstyle[prl,aps,twocolumn,epsfig]{revtex}
\draft
\begin{document}
\bibliographystyle{prsty}

\wideabs
{
\title{Shock Wave Emissions of a Sonoluminescing Bubble}
\author{Joachim Holzfuss, Matthias R\"uggeberg and Andreas Billo}
\address{Institut f\"ur Angewandte Physik, 
TU Darmstadt, Schlo\ss gartenstr. 7,
64289 Darmstadt, Germany\\
(Copyright 1998 by The American Physical Society,
Phys. Rev. Lett. {\bf 81}, No. 23 (1998))
}


\maketitle
\begin{abstract}
A single bubble in water is excited by a standing ultrasound wave.  At
high intensity the bubble starts to emit light. 
Together with the emitted light pulse, a shock wave is generated in
the liquid at collapse time.  The time-dependent velocity
of the outward-travelling shock is measured with an imaging technique. The
pressure in the shock and in the bubble is shown to
have a lower limit of 5500 bars. 
Visualization of the shock and the bubble at different
phases of the acoustic cycle reveals previously unobserved dynamics during
stable and unstable sonoluminescence.
\end {abstract}
}

When intense ultrasound sound is applied to water, bubbles appear in
the liquid. Among the properties they exhibit is sound
radiation 
and emission of photons\cite{leighton}.
In a controlled experiment, a single bubble alone may
be driven stably in a standing ultrasound field.  Here, intense light
pulses of very short duration may be observed.  Since this discovery
\cite{felipe} experimental work on the so called single bubble
sonoluminescence (SBSL) has been extensively carried out to explain the
phenomenon and the interesting features it displays:  The energy is
focussed by 12 orders of magnitude\cite{barber}, the light pulses are
of picosecond duration \cite{flash}, the
emitted light energy per pulse is in the MeV range, 
the blackbody-like spectrum peaks in the ultraviolet. The
inter pulse synchronicity can be accurate on the picosecond scale
\cite{barber} or chaotic on the microsecond scale \cite{holt94}. 
Parameter studies have been done showing 
the region of stable SBSL lying on the boundary of a dissolution island
\cite{holt96}.
Advanced driving of the bubble is employed to increase the light output
\cite{boost_prl}. 
Theoretical and numerical work \cite{th,moss,lohse} 
has been done to explain SBSL
but so far few basic assumptions of the different
theories could be verified experimentally.
An inner shock wave launched in the interior of the bubble
upon collapse has theoretically been assumed to account for the
observed short SBSL light pulse and its spectrum
\cite{hiller98,wuroberts,moss}.

\begin{figure}
\begin{center}
\epsfig{file=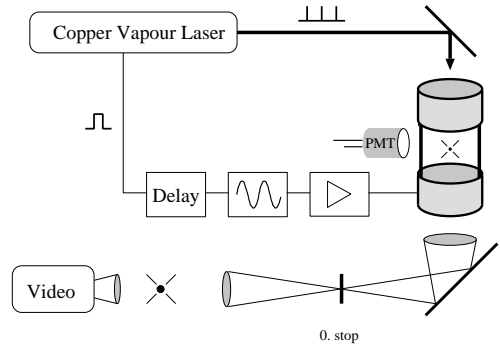,width=6.5cm}
\end{center}
\caption{Experimental setup} 
\label{setup} 
\end{figure}

Our experimental and numerical work focuses on the observation of
shock waves being emitted into the liquid at bubble collapse
\cite{firstshock}.  The shock waves are visualized, the velocity of the
front as it travels outwards is measured and the peak pressure of
the shock is deduced.  Effects appearing at unstable SBSL are
analyzed. The experiments are consistent with numerical simulations.
In the experiment (figure \ref{setup}) the standing
ultrasound wave is produced in a cylindrical cell filled with water of
ambient temperature, distilled and de-gassed to 10-40\% of ambient gas
pressure.  The cell consists of two piezoceramic cylinders connected by
a glass tube \cite{holt90} of 2.9\ cm radius (overall height 
12~cm).  An optical glass plate closes the bottom,
the top remains open.  The driving frequency is 23.5\ kHz and the driving
amplitude $\approx 1.2 \hbox{ to }1.5 \hbox{ bars}$. A bubble is
\noindent inserted into the liquid with a syringe.  The oscillating bubble
is illuminated from the top by a copper vapour laser by light 
pulses of 7\ ns duration (FWHV) followed by a low intensity tail of 30\ ns. 
The wavelength is 511\ nm.  
The repetition rate of one half the driving frequency 
is adjusted via a controllable delay to
accommodate either locking to the driving signal or 
controlled phase shifting.  
 
\begin{figure}[t]
\centerline{%
  \frame{\psfig{figure=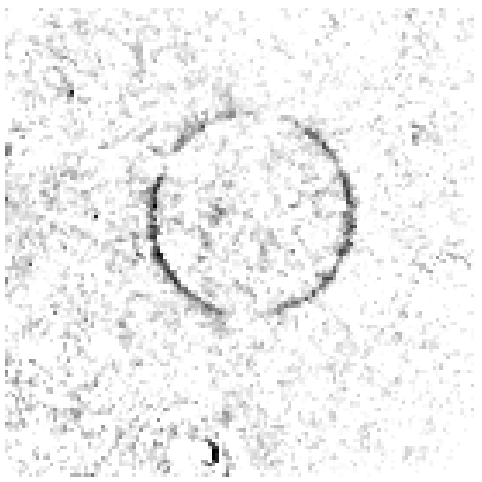,width=3.6cm}}~
  \frame{\psfig{figure=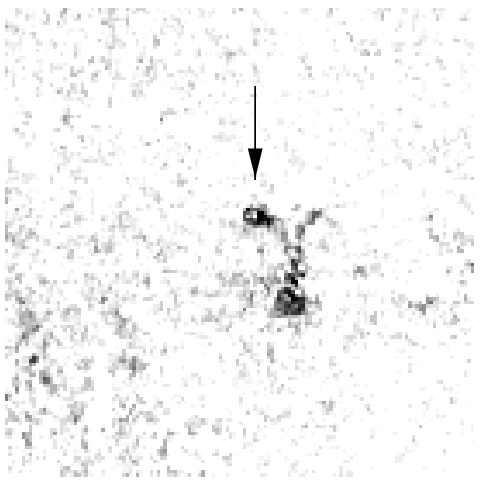,width=3.6cm}}%
}
\vspace*{1ex}
\caption{Images of shock waves emitted by a sonoluminescing bubble.
a) at $t=480\ $ns after collapse b) reflection from the side walls at
$t=3.54\ \mu$s before collapse. The arrow marks the bubble position;
image side length is 3.5\ mm.} 
\label{shock}
\end{figure}

Because shock waves modulate the phase of
the laser light, optical filtering is used to transform this
information into intensity modulations. 
Therefore the bubble image is passed through a (magnifying) $4f$ spatial
Fourier filter.  Specifically, a Dark-Ground method
\cite{HechtZajac} is used that removes the zeroth order in
the Fourier plane with a thin metal stick.  Subsequently the
image is picked up by a video camera
delivering 25 frames/s.  The shutter opening time is 0.25~ms
such that the average image of 2--3 shock waves is seen.
Because of the stable repetitive bubble
collapse a slow motion video of the oscillations \cite{holt90,tian} and
the shedding of shock waves is produced by slightly detuning
the laser flash frequency.  The images of the shock waves
are digitized in a computer and their radius/time curves can
be plotted.  Because of their submicron size, sonoluminescing bubbles
are hard to detect at collapse.  But
by recording the center of the shock wave the bubble position 
can be determined.
Figure \ref{shock} shows the images of shock waves at different times.
The shock is emitted at the main collapse of the bubble. The
shock front shows up as a circle (figure\ \ref{shock}a) and no anisotropy 
\noindent is seen within the optical resolution limit of 1.5\ $\mu$m, which is an indication 
of a symmetrical collapse. The
front proceeds  to the outer glass wall of the cylinder, reflects
and moves inward again.  The reflected shock wave has a duration $>$
40\ ns and is distorted, presumably due to imperfections or
misalignment of the glass wall. 
Figure\ \ref{shock}b shows the shock wave at the
time it is refocused the most (3.54 $\mu$s before the next collapse).
The main pressure peak seems to be $\approx 700\ \mu$m away from the bubble
at the lower end of a line structure. The refocused shock is sometimes powerful enough to
kick the bubble through space a bit as it passes it. Weaker
secondary reflections are also observable. At no time we could see a
pressure pulse due to bubble rebounds \cite{matula_shock}.
The duration of the shock pulse can be determined to be
10\ ns (FWHV). As this is on the order of the optical pulse length of 7\ ns, 
this value is an upper bound.

\begin{figure}[b]
\vspace*{-0.8cm}
\psfig{file=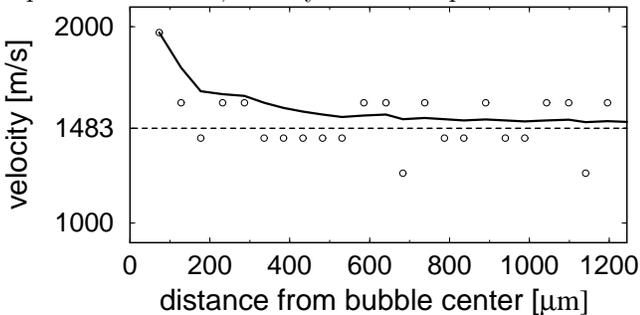,width=4cm,angle=-90}
\vspace*{0.2cm}
\caption{Average velocities ($t_{ave}=34\ ns)$ of the SBSL-shock wave
from successive images as a function of the distance from the
generation (circles).  The solid line is the mean velocity extrapolated
by averaging over all measurement points up to a respective 
distance $r$ from the bubble center.}
\label{expv}
\end{figure}

From successive images the velocity of the 
shock front is calculated. Figure\ \ref{expv} shows the average 
velocities as a function of 
distance from the bubble center. At very small distances (6-73\ $\mu$m)
an average value of $\bar v=2000$\ m/s is measured, at larger distances 
the velocity of the shock front is decreasing to
the ambient sound speed. Because the velocity of the shock 
decreases rapidly, the instantaneous velocity may be well above 2000\ m/s.
The pressure $p$ in the shock can 
be determined
from its velocity $v$ by a Rankine-Hugeniot relation \cite{morse} 
and a state equation for water, namely the Tait equation
 \begin{equation}
v={1\over \rho_0} \sqrt{
			{p-p_0} \over {
					\rho_0^{-1} - \rho^{-1}
				      }
		       } 			
~~\hbox{and}~~
  \frac{p+B}{p_0+B}={\left\{\frac{\rho}
  {\rho_{0}}\right\}}^n~~~.\\
\label{eq:tait}
\end{equation}
$\rho$ and $p$ are the maximal density and pressure in the shock,
$\rho_0$=998.2 kg/m$^3$ and $p_0$=1 bar are the ambient density and
pressure, $n$=7.025, $B$=3046 bars \cite{nist}. Using (\ref{eq:tait}), the shock 
pressure can be calculated to be 5500\ bars.

Numerical calculations have been 
carried out to further analyze the time dependence of the velocity and pressure 
of the shock front. 
The Gilmore model \cite{knapp-daily} describing the 
radial motion of a bubble is used.
\begin{eqnarray}
\left( 1 - \frac{\dot{R}}{C} \right)R\ddot{R}&+&\frac{3}{2}\left( 1
-\frac{\dot{R}}{3 C}\right) \dot{R}^2 \nonumber \\
&=& \left( 1+\frac{\dot{R}}{C}\right) H+
\left(1-\frac{\dot{R}}{C}\right)\frac{R}{C}\frac{\mbox{d}H}{\mbox{d}t}~~,
\label{eq:gilmore}
\end{eqnarray}
\begin{eqnarray}
  \left. C = c \right|_{r=R} &=&	
	\left. \sqrt{\frac{\mbox{d}p}{\mbox{d}\rho}} \right|_{r=R} = 
  c_0\left(\frac{p(R,\dot{R})+B}{p_0+B}\right)^{(n-1)/2n}, \nonumber \\
  \nonumber \\
p(R,\dot{R})&=&\left(p_{0}+\frac{2\sigma}{R_{0}}\right){\left(\frac{R_{0}^3
-a^3}{R^3-a^3}\right)}^{\kappa}
-\frac{2\sigma}{R}-\frac{4\mu}{R}\dot{R}~~.\nonumber 
\end{eqnarray}
$R$ is the bubble radius, $C$, $\rho$, and $p$ are the speed of sound in
the liquid, its density, and the pressure at the
bubble wall, respectively. $H =\int\nolimits^{p(R)}_{p_{\infty}}
{\rho^{-1}}\mbox{d}p$
is the enthalpy  of the
liquid. Parameters were set to $c_0$=1483~m/s,
$\sigma$=0.0725\ N/m, $\mu$=0.001\ Ns/m$^3$.
$a$=$R_0$/8.86 is a hard-core van der Waals-term \cite{loeff} 
and $\kappa$=5/3 the
adiabatic exponent for argon \cite{lohse}. The pressure at infinity is
$p_\infty=p_0+p_e cos(2\pi f t)$, $p_e$ and $f$ are the driving
pressure and frequency. 

The dynamics of the pressure pulse in the liquid is calculated by using the
Kirkwood-Bethe hypothesis \cite{knapp-daily}: the invariant quantity
$Y=R(H+\frac{1}{2}\dot{R}^2)$ propagates with the characteristic velocity
$c+u$, the local sound plus particle velocity
in the liquid. The outgoing characteristics are determined by \cite{lee}
\begin{eqnarray}
  \frac{\mbox{d}u}{\mbox{d}t} &=&
	\frac{1}{c-u}
	\left[(c+u)\frac{Y}{r^2}-\frac{2c^2u}{r}\right], \nonumber \\  
  \frac{\mbox{d}p}{\mbox{d}t} &=& 
	\frac{\rho_0}{r(c-u)}
	\left(\frac{p+B}{p_0+B} \right)^{1/n} 
	\left[ 2c^2u^2-\frac{c(c+u)}{r}Y \right]. \nonumber
\end{eqnarray}
Solving the bubble equation (\ref{eq:gilmore}) gives the initial values
$R$, $\dot{R}$, $H$ and $u=\dot{R}$ for each characteristic. Crossing
characteristics in $r-t$ space imply the generation of a shock. The
exact position of the shock front can be obtained by equalization of
the particle velocities in the hysteretic $u-t$ curves\cite{rudenko}.

\begin{figure}[htb]
\epsfig{file=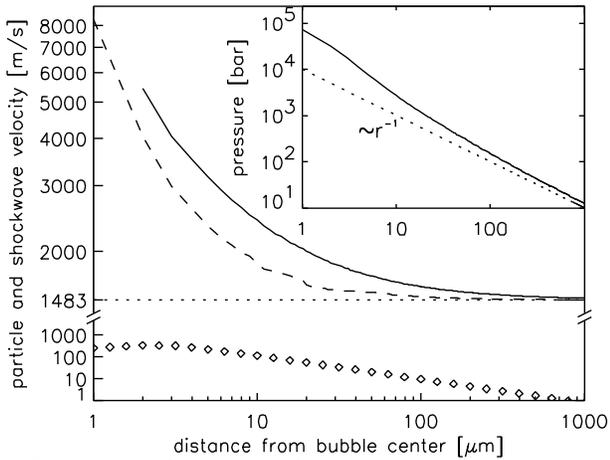,width=8cm}
\caption{Numerical calculations of a shock wave generated at collapse 
time of a bubble of 5 $\mu$m ambient radius driven 
at 23.5 kHz and 1.45 bars. Shown are data for the pressure peak that is 
travelling away from the bubble into the liquid
as a function of distance from the bubble center. 
Dashed line: velocity of the peak, 
solid line: extrapolated mean shock velocity, 
squares: particle velocity; 
the inset shows the calculated peak pressure 
in the shock as a function of distance from the bubble (solid line). 
The dotted line shows a $r^{-1}$ reference line for comparison.
}
\label{velo}
\end{figure}

Figure \ref{velo} shows the calculated shock wave velocity as a function of
the distance from the bubble center. The maximal
velocity of the pressure peak of 
\noindent approx. 8300 m/s 
decreases within the first hundred $\mu$m to the ambient sound
velocity. For comparison with the experiment the mean velocity of the peak
is shown in figure \ref{velo}. The
experimentally obtained short time average and mean velocities compare 
quite well to the numerical findings. 
The particle velocity in the model reaches a maximum value of 333 m/s. The
inset in figure \ref{velo} shows the peak pressure of the shock as it travels away
from the center. The maximum value of 73000 bars at 1 $\mu$m decays
quickly with increasing distance, within a hundred $\mu$m with a
faster decay rate than the usual $r^{-1}$. Though these numbers may be somewhat
overestimated due to model limitations,
it is seen that within the
first few $\mu$m extreme conditions exist in the fluid. 
The greater dissipation close to the bubble may account for
differences between our experimental results and 
previous inferences of the shock pressure
from direct hydrophone measurements, 
which have yielded smaller values for the pressure\cite{matula_shock,weninger}.

Using the shock wave  as a microscope for
the bubble position at collapse time, the time dependency  
and the position of the
collapse have been measured for unstable SBSL. Unstable SBSL 
occurs at the upper parameter values of the 
driving pressure and ambient gas concentration: 
The ambient bubble radius grows until the bubble dynamics reaches an 
instability where bubble volume is rapidly lost. This cycle
repeats itself on a slow time scale\cite{felipe,barber95}. Using
rare events of double exposure at split-off time, the distance of the
centers of two shock waves representing a bubble before and after the
split-off can be used to calculate a lower bound of the bubble velocity
due to the recoil of 0.5\ m/s. 
During unstable SBSL the
\begin{figure}[htb]
\center{\epsfig{file=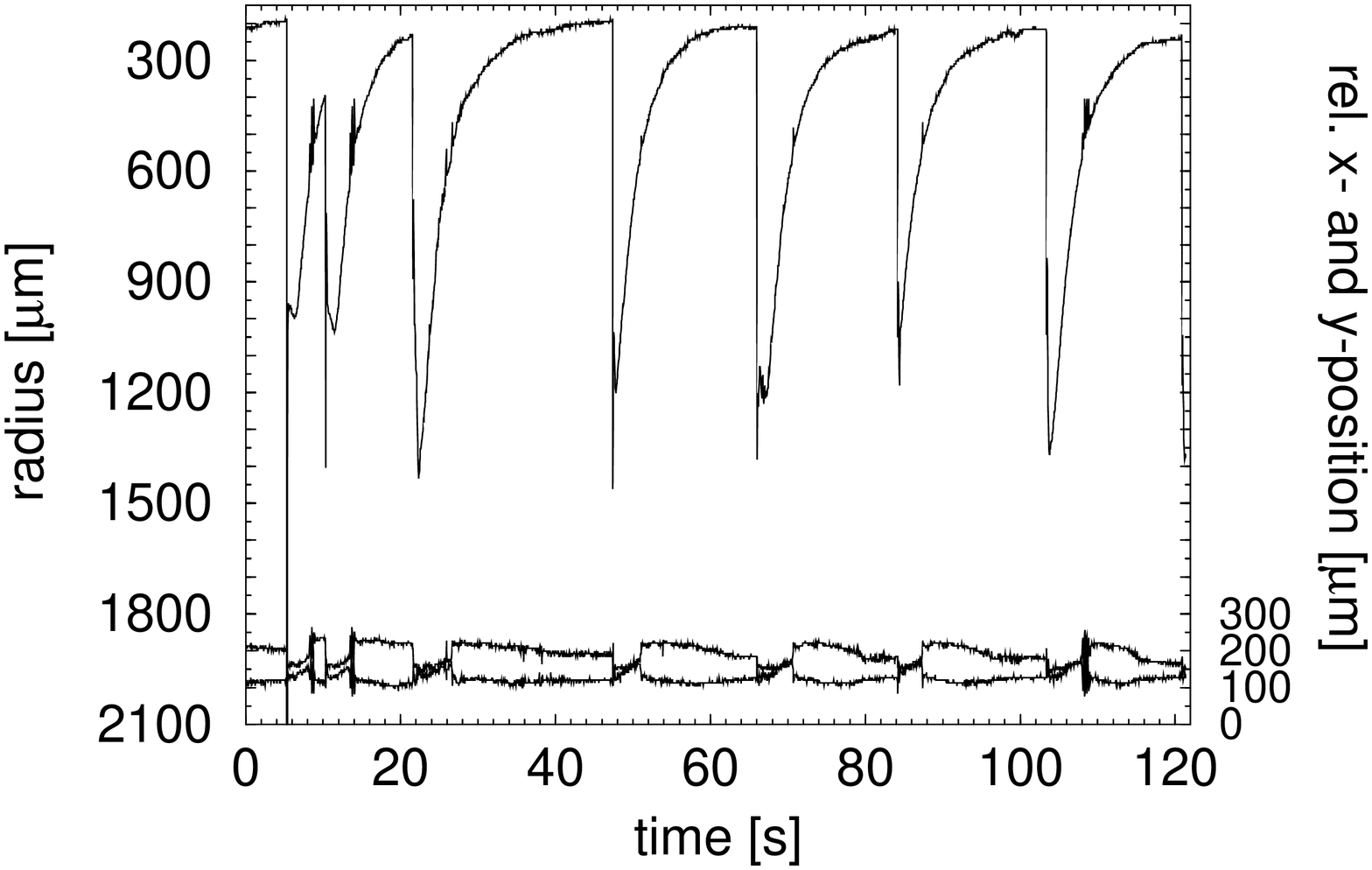,width=5cm,angle=-90}}
\center{\epsfig{file=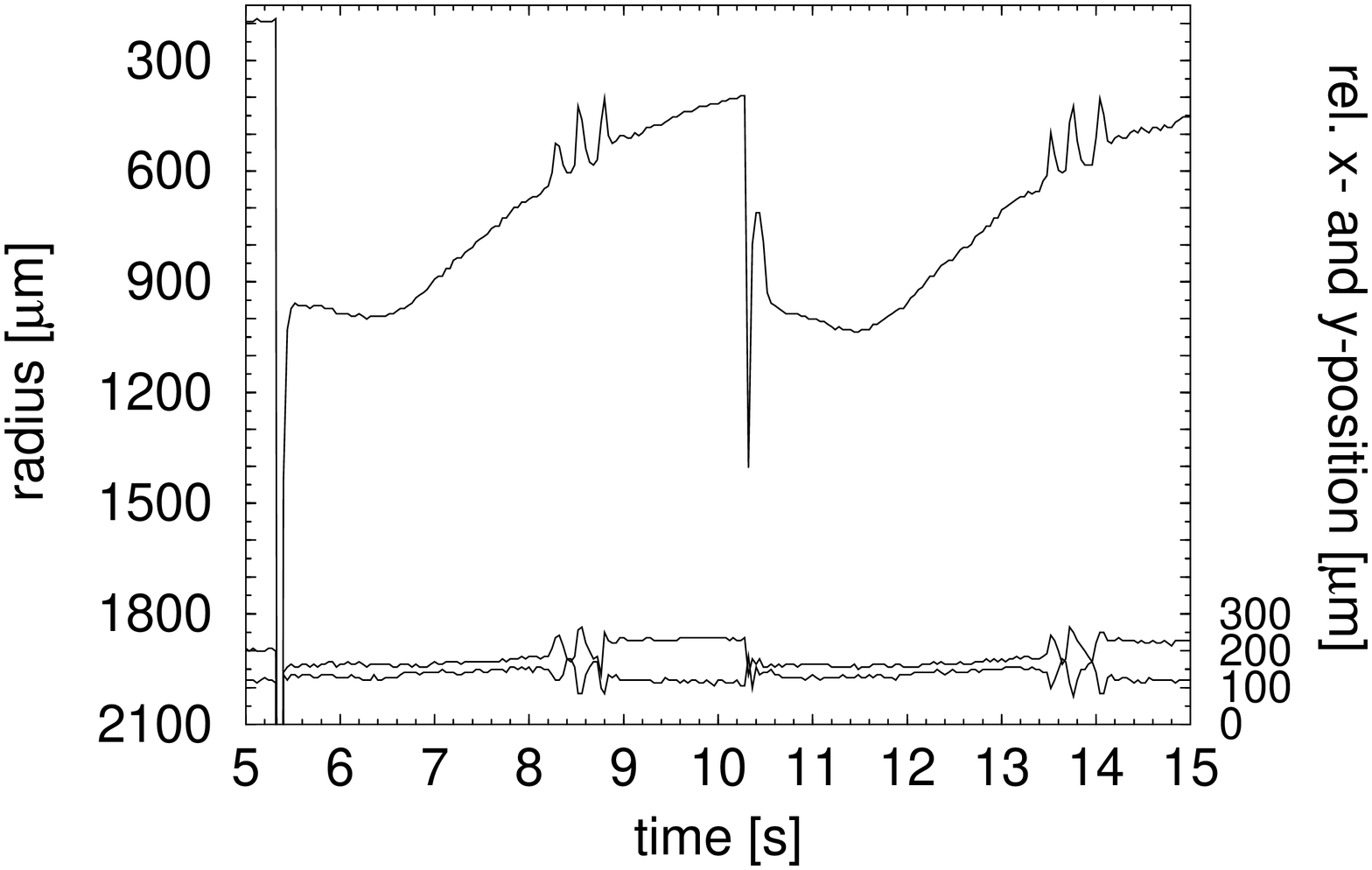,width=5cm,angle=-90}}
\caption{Measured radii of shock waves and position of the center
during unstable SBSL. Data have been digitized from images taken
each 40\ ms at a constant phase of the driving. a: The upper line
shows the radii of shock waves as a function of time, as they 
change on a slow time scale. The lower two lines are relative x- and
y-coordinates of the bubble at collapse.
b: Zoom into the first seconds of upper figure.
Spatial oscillations are seen as the shock radii 
(and the ambient radius of the bubble) are changing 
at e.g. t=8-9\ s.
}
\label{unstable}
\end{figure}
\noindent bubble shows its dynamical
behaviour over a long range of its ambient radius as a parameter. 
Figure \ref{unstable} shows the radius of the shock wave and the bubble
position at collapse time as a function of time. All experimental 
conditions were kept constant.  It is seen that the bubble collapse does
not occur at a constant phase any more.
As the ambient bubble radius grows by diffusion, the
collapse is shifted to later times. Because the illuminating flash 
occurs at a fixed phase of the driving signal some small time 
after the collapse, a later collapse decreases the time the shock
front can travel outward until it is imaged.
This way a larger
ambient bubble radius shows up as a smaller shock radius.
In  figure \ref{unstable}a the recurrent process of growing
on a slow time scale and a subsequent rapid decrease of ambient volume of the 
bubble is seen. 
At split-off the collapse time of the bubble is shifted by $\approx 1~\mu$s with
respect to the driving phase. Calculations of collapse time vs. bubble
volume for SBSL-relevant bubble radii show (see also \cite{felipe}) 
that the bubble loses about one half of its volume.
Most probably, multiple fragments (micro bubbles) will be generated.
So far micro bubbles have been observed at
the lower amplitude threshold of SBSL, where they have a slowly
decreasing velocity of initially 3-4 mm/s, 
do not return to the bigger bubble but dissolve within a few
hundred $\mu$m within $\approx 0.1$\ s.
A closer look on a single cycle of figure \ref{unstable} reveals, that
the growing phase has a peculiar fine structure. Immediately after 
micro bubble split-off (e.g. at t=5.4\ s, 10.4\ s) 
the collapse is shifted to later times (smaller shock radii), 
reaching a slightly
decreasing plateau until it finally increases.
During each growing phase
small bumps are seen (e.g. at $t$=8--9\ s). 
Looking at the position of the 
bubble one sees a connection: Each time the phase bumps, the bubble 
moves discretely through space and finally settles.
The explanation may be oscillation 
in different resonances, shock wave interaction or 
the acoustic field acting on the bubble is altered as it grows.

We have visualized the generation of shock waves from a sonoluminescing
bubble for the first time.  Resulting from the enormous pressure inside
the bubble and the great amount of energy transported by the
surrounding liquid, the shock front is shown to have a faster speed
than the ambient sound velocity.  
If one calculates \cite{refrac} the change in the refractive index 
of water due to the
theoretical local overpressure of 73 kbars, one arrives at a
$\Delta n=0.23$, which is 70\% of the change at an air/water interface. 
Therefore measurements of the minimal radius by Mie scattering
together with statements about the exact timing of the flash with
respect to the minimal radius should be done keeping this in
mind as the shock wave builds a scattering layer around the bubble. 
We cannot conclude from our data,
whether the visualized shock in the liquid also consists of
contributions of an hypothetical inner bubble shock that may be
responsible for SBSL.  The observed refocused shock can be shown to
have an impact on the bubble dynamics.  It would be interesting to see,
if exact positioning and control of the timing of the reflected 
shock wave can be used to increase SBSL
intensity \cite{boost_prl,moss}.

The authors thank 
R.\ G.\ Holt for continuous collaboration, R. A. Roy and T. J. Matula
for discussions,  S. J. Putterman for stimulating questions and
the TU Darmstadt for making 
the research possible.  The work has been
funded through the SFB 185 ``Nichtlineare Dynamik'' of the DFG.


\vspace*{-0.5cm}
\begin{thebibliography}{99}
\vspace*{-1.5cm}
\bibitem{leighton}T. G. Leighton, {\it The Acoustic Bubble}, 
	(Academic Press, London, 1994);
	W. Lauterborn and J. Holzfuss, 
	J. Bif. and Chaos {\bf 1}, 13 (1991).
\bibitem{felipe}D. F. Gaitan, L. A. Crum, C. C. Church, and R. A. Roy,
         J. Acoust. Soc. Am. {\bf 91}, 3166 (1992).
\bibitem{barber}B. P. Barber, R. A. Hiller, R. Löffstedt, 
	S. J. Putterman, and K. R. Weninger, Phys. Rep. 281, 65 (1997).
\bibitem{flash}
	B. P. Barber, S. J. Putterman, Nature {\bf 352},
        318 (1991);
	B. Gompf, R. G\"unther, G. Nick, R. Pecha, and W. Eisenmenger,
	Phys. Rev. Lett. {\bf 79}, 1405 (1997);
	R. A. Hiller, S. P. Putterman, and K. R. Weninger,
	Phys. Rev. Lett. {\bf 80}, 1090 (1998);
	M. J. Moran and D. Sweider, 
	Phys. Rev. Lett. {\bf 80}, 4987 (1998). 
\bibitem{holt94} R. G. Holt, D. F. Gaitan, A. A. Atchley, and J. Holzfuss, 
	 Phys. Rev. Lett. {\bf 72 },  1376 (1994).
\bibitem{holt96}R. G. Holt and D. F. Gaitan, 
	Phys. Rev. Lett. {\bf 77}, 3791 (1996).
\bibitem{boost_prl}J. Holzfuss, M. R\"uggeberg and R. Mettin, 
	Phys. Rev. Lett. {\bf 81}, 1961 (1998).
\bibitem{th}
	L. Fromhold, Phys. Rev. Lett. {\bf 73}, 2883 (1994);
\bibitem{moss}W. C. Moss, D. B. Clarke, J. W. White, and D. A. Young, 
	Phys. Lett. A {\bf 211}, 69 (1996);
	W. C. Moss, D. B. Clarke, and D. A. Young, Science 
	{\bf 276}, 1398 (1997).
\bibitem{lohse}
	A chemical dissociation theory states that only argon 
	remains within sonoluminescing air bubbles. See
	D. Lohse, M. Brenner, T. Dupont, S. Hilgenfeldt, and B. Johnston,
	Phys. Rev. Lett. 78, 1359 (1997) for the theory and 
	T. J. Matula and L. A. Crum, Phys. Rev. Lett. {\bf 80}, 865 (1998)
	for an indirect experimental evidence, (also \cite{holt96}).   
\bibitem{hiller98} R. A. Hiller, S. J. Putterman, and K. R. Weninger,
	Phys. Rev. Lett. {\bf 80}, 1090 (1998).
\bibitem{wuroberts}C. C. Wu, and P. H. Roberts, 
	Phys. Rev. Lett. {\bf 70}, 3424 (1993).
\bibitem{firstshock} J. Holzfuss, M. R\"uggeberg, and A. Billo, 
        Fortschritte der Akustik - DAGA 97, Bad Honnef:
        DPG GmbH (1997), pp. 341--342.  	
\bibitem{holt90}R. G. Holt, J. Holzfuss, A. Judt, A. Phillip,  and
         S. Horsburgh, in: Frontiers of Nonlinear Acoustics: 
         Proc. of the 12$^{th}$ ISNA  Eds.: M.F.
         Hamilton and D.T. Blackstock, Elsevier Science Publ.
         Ltd., London, (1990) pp. 497-502.	
\bibitem{HechtZajac}E. Hecht, {\it Optics}, (Addison-Wesley, Reading, 1987).
\bibitem{tian}Y. Tian, J. A. Ketterling, and R. E. Apfel, 
	J. Acoust. Soc. Am. {\bf 100}, 3976 (1996).
\bibitem{matula_shock}T. J. Matula, I. M. Hallaj, R. O. Cleveland, L. A. Crum, 
	W. C. Moss, and R. A. Roy, J. Acoust. Soc. Am. {\bf 103}, 1377 (1998).
\bibitem{morse}P. M. Morse and K. U. Ingard,{\it Theoretical Acoustics},
	(Princeton University Press, Princeton, 1986).
\bibitem{nist}The constants $n$ and $B$ stem from fits of the $C/p$ relation 
(in eq. (\ref{eq:gilmore})) to data from: E.W. Lemmon, 
	M.O. McLinden and D.G. Friend, 
 	in:
        {\it NIST Chemistry WebBook}, NIST Stand. Ref. Database Number 69, 
	edited by: W.G. Mallard and P.J. Linstrom, March 1998, 
	NIST,
        Gaithersburg MD, 20899 (http://webbook.nist.gov). 	
\bibitem{knapp-daily}R. T. Knapp, J. W. Daily, F. G. Hammitt,
         {\it Cavitation} (McGraw-Hill, New York, 1970). 
\bibitem{loeff} R. L\"offstedt, B. P. Barber and S. J, Putterman, 
	Phys. Fluids A, {\bf 5} 2911 (1993).
\bibitem{lee}Y.-P. Lee, S. W. Karng, J.-S. Jeon, and H.-Y. Kwak,
	J. Phys. Soc. Japan, {\bf 66}, 791 (1997).
\bibitem{rudenko}O. V. Rudenko, S. I. Soluyan,
         {\it Theoretical Foundations of Nonlinear Acoustics}
         (Consultants Bureau, New York, 1977) p.~28. 
\bibitem{weninger}K. R. Weninger, B. P. Barber, and S. J. Putterman,
	Phys. Rev. Lett. {\bf 78}, 1799 (1997).
\bibitem{barber95}B. P. Barber, K. Weninger, R. L\"ofstadt, and S. Putterman,
	Phys. Rev. Lett. {\bf 74}, 5276 (1995).
\bibitem{refrac}$({n_{r}^2-1) / ({n_{r}^2+2}})\cdot\rho^{-1}	=
		0.2060\cdot10^{-3}~\hbox{m}^3/\hbox{kg}$ 
		and Eq. (\ref{eq:tait}),
	from: H. Schardin, Erg. exakt. Naturwiss. {\bf 20}, 370 (1942).




	  
\end{thebibliography}
\end{document}